\documentclass[journal=apchd5,manuscript=article]{achemso}
\usepackage[version=3]{mhchem} % Formula subscripts using \ce{}
\usepackage{float}
\usepackage{longtable} %Per avere migliori tabelle
\usepackage{multirow}

\usepackage{hyperref}

\usepackage{pdfpages}

\usepackage[dvipsnames]{xcolor}     
\usepackage[most]{tcolorbox}

\usepackage{soul}

\usepackage{comment}

\author{Tommaso Cenci}
\author{Riccardo Alessandro}
\author{Enrico Ronca}
\email{enrico.ronca@unipg.it}
\affiliation[UniPG]
{Dipartimento di Chimica, Biologia e Biotecnologie, Universit{\`a} degli Studi di Perugia, 06123, Perugia, Italy}

\title[An \textsf{achemso} demo]
  {Toward a local manipulation of DNA by quantum fields}

\begin{document}

%%\begin{tocentry}

%Some journals require a graphical entry for the Table of Contents.
%This should be laid out ``print ready'' so that the sizing of the

%Inside the \texttt{tocentry} environment, the font used is Helvetica
%8\,pt, as required by \emph{Journal of the American Chemical

%The surrounding frame is 9\,cm by 3.5\,cm, which is the maximum
%permitted for  \emph{Journal of the American Chemical Society}
%graphical table of content entries. The box will not resize if the
%content is too big: instead it will overflow the edge of the box.

%This box and the associated title will always be printed on a
%separate page at the end of the document.

%%\end{tocentry}

\begin{@twocolumnfalse}
\begin{abstract}
  Hydrogen bonds are the fundamental backbone for deoxyribo-nucleic acid (DNA) stability. 
  %We were inspired by previous studies on the behaviour of the H$_2$O hydrogen bonds in polaritonic cavities, which showed a significant destabilizing effect. From these findings, we studied the effect the quantum field has on the hydrogen bonds found between the nucleotide bases' dimers found in DNA. We found a varying effect of the field depending on the orientation of the dimers in the cavity, with some showing stabilizing effects, others gave destabilizing effects. We propose a way to employ the cavity as a wedge for genome-editing techniques.
  In this letter we propose a new strategy based on plasmonic cavities to perform a local manipulation of hydrogen bonds in DNA. 
  The analysis is performed using state-of-the-art Quantum Electrodynamics Coupled Cluster calculations (QED-CC). We demonstrate that in standard strong coupling regimes, small but appreciable local modifications of the nucleotide bases' interactions can be induced in a totally non-intrusive manner. 
  The effect can eventually be enhanced if ultra-strong coupling conditions can be reached. 
  Our strategy provides an alternative approach to methodologies based on a collective coupling to perform optical DNA manipulation.  

\end{abstract}
\end{@twocolumnfalse}

%\section{Introduction}

Genetic engineering has been one of the holy grails in biochemistry since its discovery due to its potentially revolutionary applications in many fields ranging from medicine, pharmacy up to industry and agriculture ~\cite{pena1994paternity,wang2012rapid,mittler2010genetic,favier2015paraganglioma,porteus2019new,kang2021chloroplast}.

For these reasons, in the past decade, an exponential increase in both the private and public investments dedicated to this sector has been observed and a significant growth has been also foreseen for the near future~\cite{wong2023estimated}.
The attention to this field grew also thanks to the advent of accurate experimental techniques to apply local manipulations of deoxyribo-nucleic acid (DNA) strands. ~\cite{cui2018review,grunenwald2003optimization,newton1997pcr,valasek2005power} Among these techniques, the clustered regularly interspaced short palindromic repeats (CRISPR) approach has revolutionized our capabilities in doing genetic modifications both in vitro and in vivo. ~\cite{zhan2019crispr,pulecio2017crispr,wang2016crispr,li2023crispr,doudna2014new}
The double helix structure of DNA is composed of two antiparallel strands of nucleotides connected by deoxyribose and phosphate groups. 
Nucleotides pair specifically (adenine with thymine and cytosine with guanine) via hydrogen bonds, representing the main forces that maintain the stability of the whole helicoidal structure ~\cite{poater2014b,mo2006probing,goodman1997hydrogen,kool2001hydrogen} .
The CRISPR technique uses the insertion of a guide RNA (gRNA) strand in a specific region of the gene. The gRNA activates a nuclease enzyme (i.e., Cas9) able to break double-strands in the selected region. At this moment, the cell activates some repair mechanisms (nonhomologous end joining—NHEJ or homology-directed recombination—HDR) responsible for the actual genomic manipulation ~\cite{morrical2015dna,tang2019methods,liao2024crispr,su2016crispr,geisinger2016vivo,guo2018harnessing,yan2020crispr} . 
Emmanuelle Charpentier and Jennifer Anne Doudna to have reproduced the CRISPR mechanism in a lab were appointed in 2020 with the Nobel Prize in Chemistry.~\cite{doudna2014new}
%%At the moment CRISPR is the reference technique for gene editing due to its high versatility at a relatively low cost as well as the possibility to be applied both ex and in vivo.\revER{(CITA)}
Despite these advantages, the CRISPR technique suffers from possible off-target or unexpected modifications, limiting its use in several fields of application ~\cite{cho2014analysis,zhang2015off,guo2023off} .
For all these reasons, researchers are still dedicating large efforts to formulate novel approaches for DNA manipulation that are able to edit genes in a more precise but, at the same time, less intrusive manner. 

In the past decade polaritonic chemistry became very popular in the scientific community as a new tool to manipulate the properties of matter by means of quantum fields confined in optical cavities. ~\cite{ebbesen2016hybrid, herrera2016cavity, galego2017many, martinez2018polariton}
Inside an optical cavity (i.e., Fabry P{\'e}rot resonator), the states of the matter couple with the states of the field, generating new hybrid states called polaritons.
Polaritons usually have characteristics that can be very different from the bare matter states, offering an unprecedented opportunity to control the properties of matter in a totally non-intrusive manner. 
The popularity of the polaritonic chemistry field increased enormously thanks to the studies performed by the Ebbesen group, who proposed quantum fields as a new tool to manipulate chemical reactivity. In particular, they demonstrated that Fabry-P{\'e}rot cavities set in resonance with specific electronic or vibrational transitions of the reagents can be used to control the chemistry or photochemistry of a reaction. ~\cite{thomas2016ground, lather2019cavity, thomas2019tilting, delpo2020polariton, polak2020manipulating, sau2021modifying, ahn2023modification}
Recently, the Ebbesen group proposed polaritonics as a new technique to destabilize the hydrogen bonds in DNA and induce its melting (opening of the double helix).~\cite{tao2025probing}
Before Ebbesen's group, the idea of using vibrational strong coupling (VSC) to manipulate biological systems' properties with a focus on DNA was already proposed by Zhang's group. 
In particular, they also observed effects induced by quantum fields on the melting temperature of DNA, attributing the effects to a cooperative VSC with the solvent molecules. ~\cite{zhang2023mid, gu2023regulation, li2023thz, zhong2023driving, hou2024vibrational}
In both Ebbesen's and Zhang's experiments, strong coupling is reached by collective effects (coupling of many emitters to the cavity field), requiring the use of a macroscopic sample. 
In these conditions the cavity field has effects on the whole DNA chain, introducing limitations to a local control of the modification. 

\begin{figure}[H]
    \centering
    \includegraphics[width=0.6\linewidth]{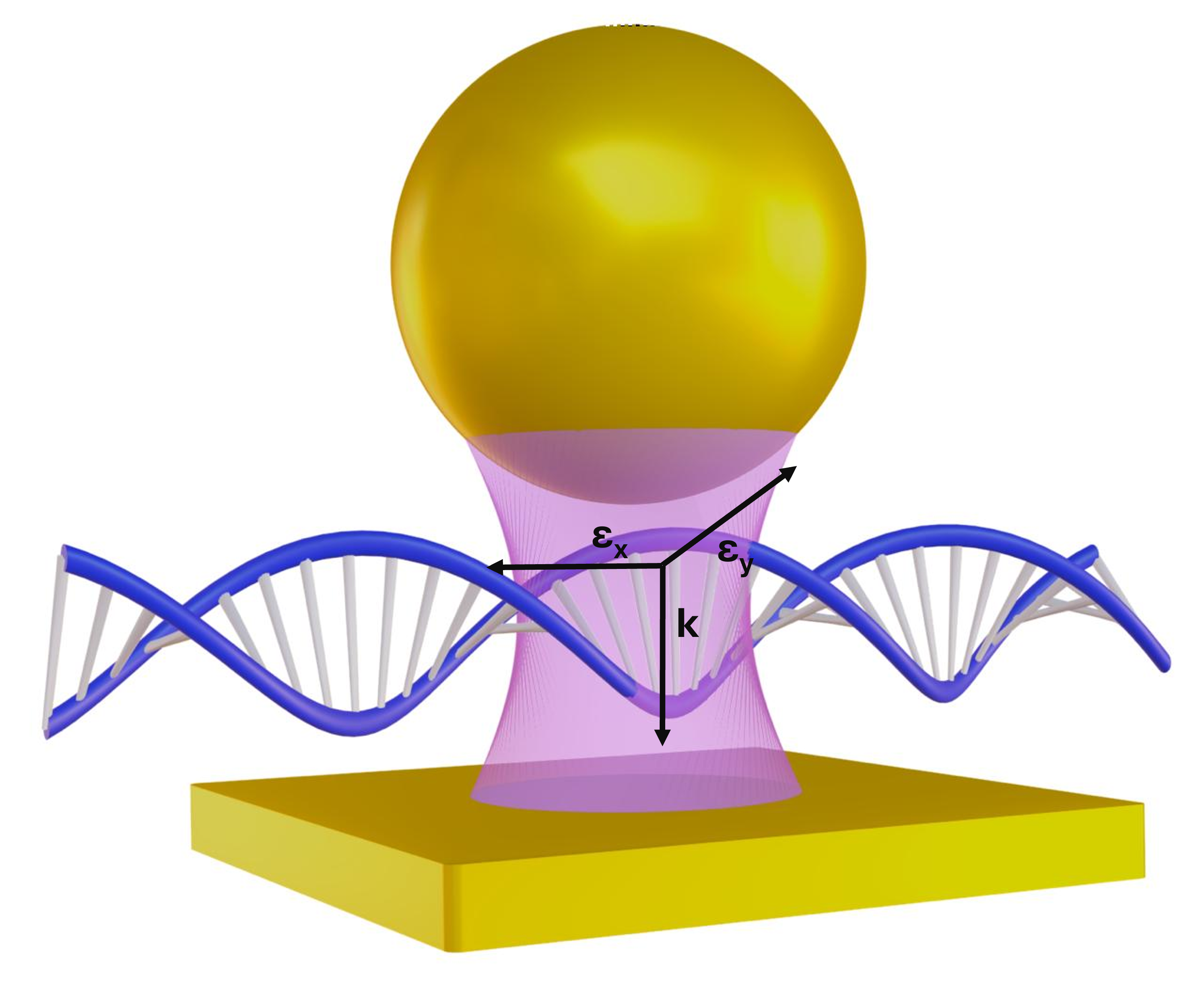}
    \caption{Pictorial representation of a DNA double strand in a plasmonic cavity.}
    \label{fig:Fig1}
\end{figure}

In this paper, we propose a new strategy based on plasmonic cavities to perform local (basis-by-basis) manipulation of a DNA double helix. 
Inside a plasmonic cavity (see Fig.~\ref{fig:Fig1}) very large coupling values can be reached also for a single (or for a very reduced number) of emitters. ~\cite{vu2025modeling, chikkaraddy2016single, baumberg2022picocavities} 
In this way, manipulation at the level of the single or few nucleotides pairs is in principle possible. 

The analysis has been performed by means of theoretical \textit{ab-initio} simulations. In order to treat the matter and the field on an equal footing, quantum electrodynamics (QED) extension of quantum chemistry approaches need to be used. 
In this study, the quantum electrodynamics coupled cluster (QED-CC) theory has been applied. ~\cite{haugland2020coupled, mordovina2020polaritonic}
QED-CC is, since its first formulation,~\cite{haugland2020coupled} the reference \textit{ab-initio} technique to investigate molecular systems strongly coupled to quantum fields. 
Even though the majority of the established \textit{ab-initio} quantum chemistry approaches (i.e. HF, ~\cite{riso2022molecular, el2024toward, castagnola2025strong} DFT,~\cite{ruggenthaler2014quantum, ruggenthaler2018quantum, flick2018ab} MP2 ~\cite{el2025strong}, CASSCF ~\cite{alessandro2025complete, vu2025modeling}, DMRG ~\cite{matousek2024polaritonic}, FCI, ~\cite{haugland2020coupled} etc.) have been extended to QED, coupled cluster is still by far the best compromise between efficiency and accuracy, especially for molecular systems involving weak interactions (i.e., hydrogen bonds, van der Waals, etc.). 
QED-CC theory has already been used by our group to predict cavity-induced modifications of intermolecular interactions. ~\cite{haugland2021intermolecular} 
In particular, we demonstrated that for this kind of systems, an actual polarization-dependent modification of the binding energy can be obtained if sufficiently large coupling values are reached. 
Theoretically, the potential energy surface reshaping is only visible if correlation effects are included explicitly. 
This observation justifies the use of QED-CC theory for our investigations of DNA.

\paragraph{Results and discussion.}

Considering the small dimensions of the plasmonic cavity, our investigations have been performed on single nucleotide bases dimers (adenine-thymine -- AT and cytosine-guanine -- CG) coupled to the electromagnetic field. This approximation can be considered valid since only a single couple (or at most just a few of them) will interact with the field confined inside the device.

The geometries of the dimers (see Supplementary Materials) have been optimized with the ORCA software package ~\cite{ORCA, neese2020orca} at the MP2 level of theory ~\cite{altun2021unveiling} using a cc-pVDZ basis set. ~\cite{dunning1989gaussian,ColucciACSCentralScience2026}

A coupling value of $\lambda=0.05$ a.u., achievable only in plasmonic cavities for a single emitter,~\cite{vu2025modeling,chikkaraddy2016single,baumberg2022picocavities} has been used throughout this paper.

For all calculations the single cavity mode approximation has been used.

Inside the plasmonic cavity, the DNA double helix can only lie perpendicular to the wave vector ($k$ -- set along the $z$ axis) of the quantum field as depicted in Fig.~\ref{fig:Fig1}. This implies that only field polarizations ($\epsilon$) on the $xy$-plane are accessible.

All calculations have been performed using the e$^\mathcal{T}$ software package.~\cite{folkestad20201, folkestad20252}

\begin{figure}[ht!]
    \centering
    \includegraphics[trim={2in 1.5in 3in 1in}, clip, width=0.6\linewidth]{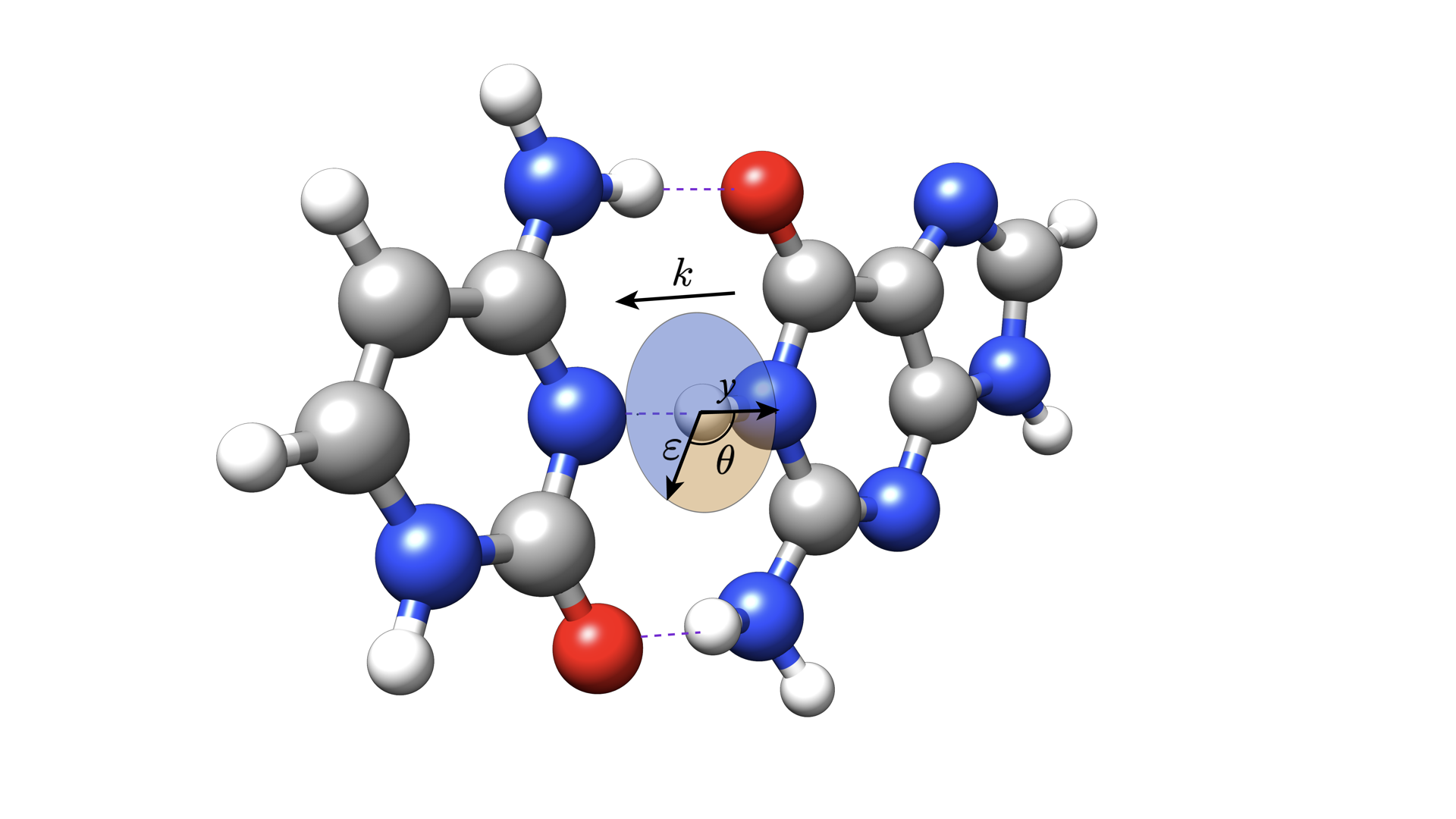}
    \caption{%Representation of the $\theta$ angle between the polarized field and the hydrogen bonds
    Graphical representation of the angle ($\theta$) defining the polarization direction.}
    \label{fig:Fig2}
\end{figure}

As a first step we investigated the effects induced by the field on the hydrogen bonds dissociation potential energy surface as a function of the polarization direction.  
The analysis has been performed by varying the angle ($\theta$) defining the orientation on the $xy$-plane of the field polarization vector ($\epsilon$) with respect to the $y$ axis (see Fig.~\ref{fig:Fig2}).

\begin{figure*}[ht!]
\includegraphics[width=\textwidth]{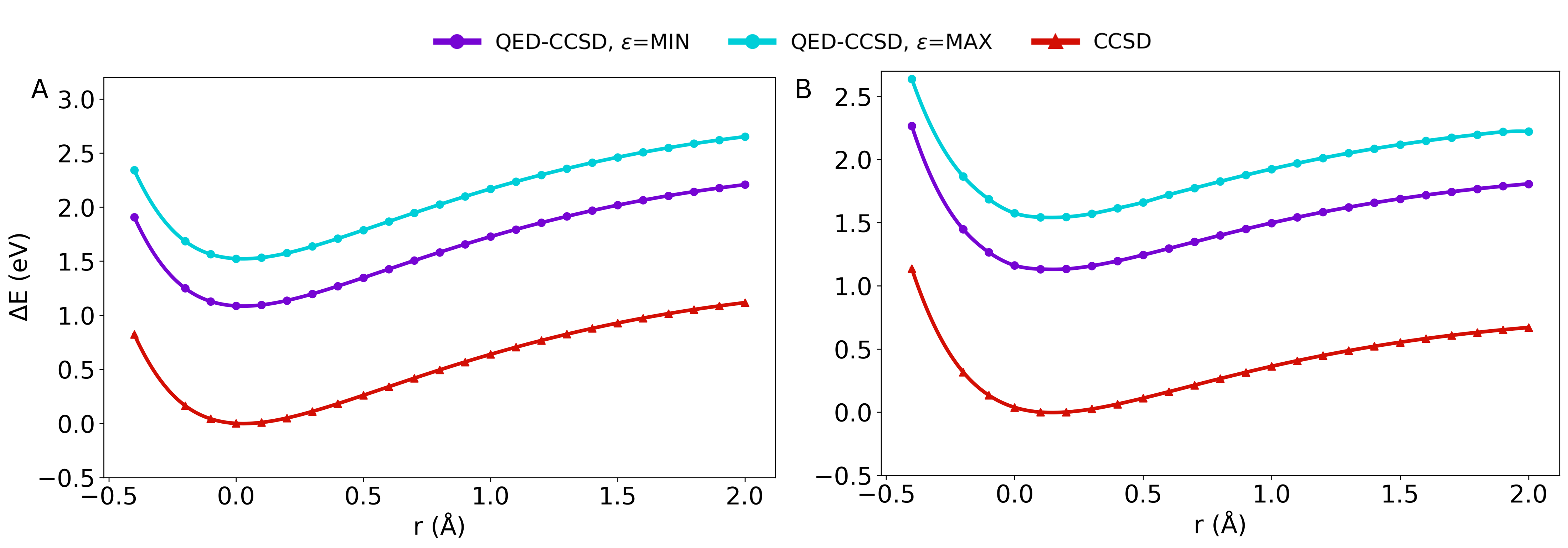}
\caption{H-bond potential energy curves for the CG (Panel A) and AT (Panel B) dimers outside (red) and inside the cavity. Cyan curves refer to the MAX polarization while the purple ones to the MIN polarization.}
\label{fig:Fig3}
\end{figure*}

Potential energy surfaces for the hydrogen bonds dissociation outside and inside the cavity with respect to different polarization directions are shown in Fig.~\ref{fig:Fig3}.
%The selected polarization directions  have been chosen for both dimers in situations when the cavity field effects is minimum (MIN -- $\theta=0^{\circ}$/$\epsilon=(0,1,0)$ for CG and $\theta=165^{\circ}$/$\epsilon=(0.26,-0.97,0)$ for AT) and when it is maximum (MAX -- $\theta=90^{\circ}$/$\epsilon=(1,0,0)$ for CG and $\theta=75^{\circ}$/$\epsilon=(0.97,0.26,0)$ for AT) as estimated by the scan reported in Figure S1.
The selected polarization directions  have been chosen for both dimers in situations when the cavity field effects is minimum (MIN)  and when it is maximum (MAX) as estimated by the scan reported in Figure S1.
Details about the MIN and MAX polarization directions are reported in Table~\ref{tab:Tab1} for both the nucleotide dimers.

\begin{table}[h]
\centering
\begin{tabular}{cc|c|c}
% Row 1
\multicolumn{1}{c}{} & \multicolumn{1}{c}{} & $\theta (^{\circ})$ & $\epsilon$ coordinates \\ \hline\hline

% Rows 2 and 3 (MIN merged)
\multirow{2}{*}{MIN} & CG & 0 & (0,1,0) \\
                     & AT & 165 & (0.97,0.26,0) \\ \hline

% Rows 4 and 5 (MAX merged)
\multirow{2}{*}{MAX} & CG & 90 & (0.26,-0.97,0) \\
                     & AT & 75 & (1,0,0) \\ 
\end{tabular}
\caption{MIN and MAX polarizations for the CG and AT dimers.}
\label{tab:Tab1}
\end{table}

As a first effect we notice that, in general, the presence of the field induces an absolute destabilization of the potential energy surfaces. 
In particular, at the minimum of the potential energy surface, the field-induced energy destabilization is about 1.1 eV and 1.5 eV for the polarization along the MIN and MAX directions respectively. Similar effects are observed for both adducts.

\begin{figure*}[ht!]
    \centering
    \includegraphics[width=1\linewidth]{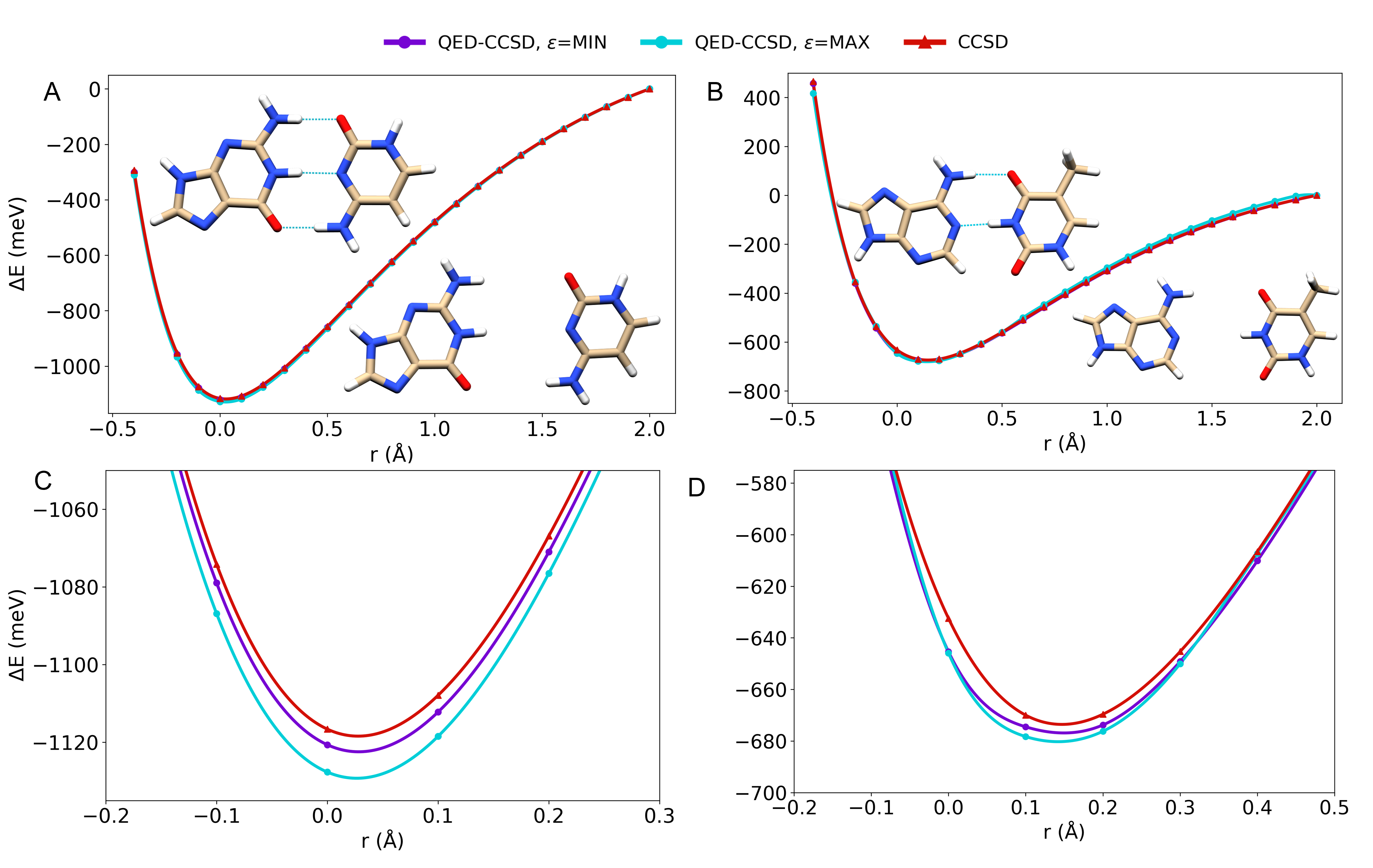}
    \caption{H-bond potential energy curves relative to separated fragments for CG (panel A) and AT (panel B). Panel C (CG) and D (AT) are zooms around the equilibrium distance to better visualize the cavity effects on the two dimers.}
    \label{fig:Fig4}
\end{figure*}

A deeper analysis (see Fig.~\ref{fig:Fig4}) highlights that the cavity field, as already observed in Ref.~~\citenum{haugland2021intermolecular} for a water dimer, can induce also actual modifications of the hydrogen bond energy.

This effect can be clearly observed in panels A and B of Fig.~\ref{fig:Fig4} respectively for the CG and AT dimers and highlighted by the corresponding zooms at the energy minima in panels C and D. 
In particular, differently from the absolute energies, we notice that the field induces a bond stabilization of about 12.4 meV ($\sim$1.1\% of the total binding energy) for the CG adduct and 13.5 meV ($\sim$2.1\% of the total binding energy) for the AT dimer estimated at the corresponding MAX polarization directions.

\begin{figure*}[ht!]
    \centering
    \includegraphics[width=\textwidth]{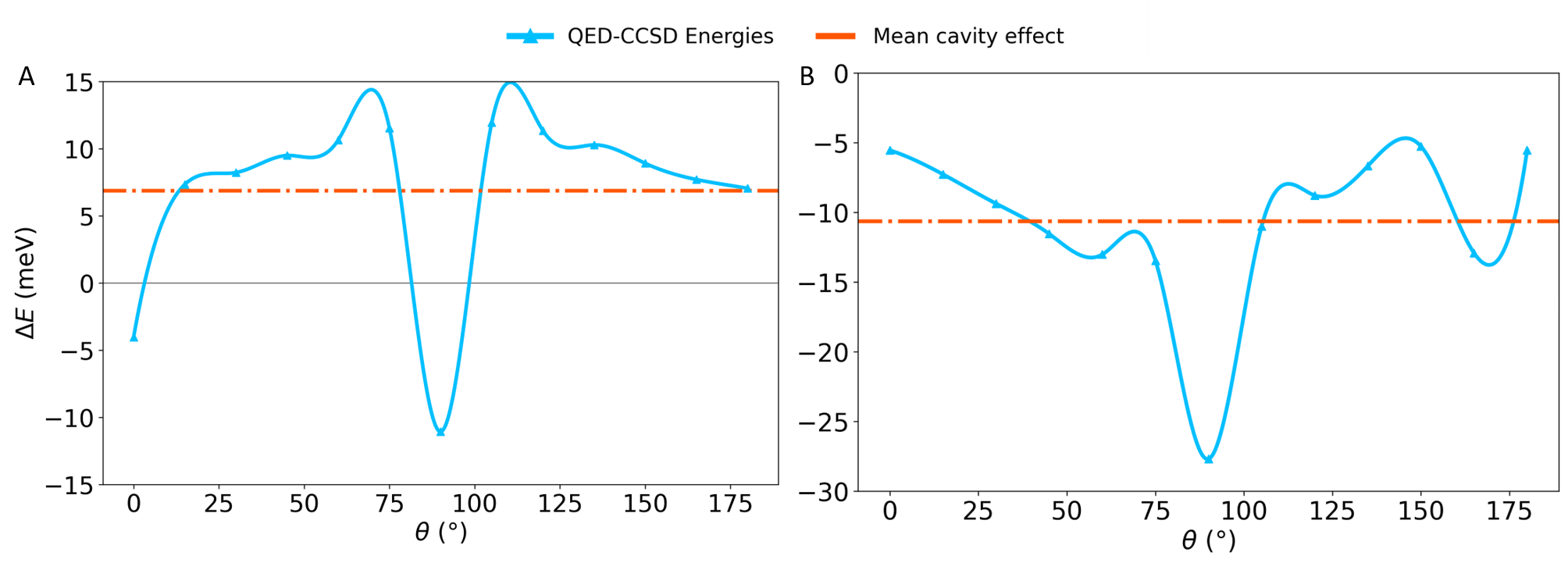}
    \caption{Cavity effects on the binding energies as a function of the field polarization direction $\theta$. Panel A refers to CG while panel B to AT. In Panel A a black horizontal line delimits field-induced stabilizations (positive) from destabilizations (negative).}
    \label{fig:Fig5}
\end{figure*}

The field-induced hydrogen bonds stabilization observed in Fig.~\ref{fig:Fig4} is also different compared to what already observed in Ref.~~\citenum{haugland2021intermolecular} for the water dimer. Indeed, in that case, the cavity field was inducing a destabilization of the binding energy. 
This effect is due to the different alignment between the field polarization and the hydrogen bond.
A similar trend was already observed in Ref.~~\citenum{haugland2021intermolecular} for 
Van der Waals interactions in the H$_2$ dimer.

To get a deeper insight on this effect, we investigated the field-induced effects on the binding energy as a function of the field polarization direction (or equivalently on the dimer orientation). The results are presented in Fig.~\ref{fig:Fig5}.

A different trend can be observed in this case for the two systems. In particular, the CG dimer (see Fig.~\ref{fig:Fig5}A) shows both stabilizing (around 0° and 90°) and destabilizing effects (elsewhere). On the other hand, the AT dimer (see Fig.~\ref{fig:Fig5}B) shows
stabilizations for every orientation of the polarization vector. However, it is important to remark that, also in the CG dimer case, the average field effect (orange dashed line) is different from zero. This ensures that the effect does not cancel out in an
eventual experiment. 
For this dimer, the field produces an average destabilizing effect of about 6.7 meV (0.6\% of the total binding energy -- -1117 meV). 

In the AT dimer case instead the average field effect induces a stabilization of the hydrogen bonds resulting in a negative mean cavity effect of about 10.6 meV, 1.68 \% of the total interaction energy (-632.4 meV).
The maximum stabilization, in this case, is about -27.7 meV corresponding to 4.3$\%$ of the total binding energy.
Due to the significantly lower binding energy, the cavity effect is, in percentage, more pronounced for the AT dimer compared to the CG one. However, it still remains quite reduced, at least for this range of coupling values.

The field-induced effects observed on the energies are associated with corresponding modifications of the electron density of the system. Density differences for the system in the cavity compared to the one outside are presented in Fig.~S2.
%\begin{figure}[ht!]
%    \centering
%    \includegraphics[width=\linewidth]{Images/DIFFDIFF.png}
%    \caption{Differences between the electronic densities of the dimers in cavity with $\theta$ = 90 and the sum of the electronic densities of the single monomers in cavity at the same geometry. Zones depicted in red have a depletion of electron density, in blue an accumulation of it. Panel A refers to the CG dimer, panel B to the AT dimer.}
%    \label{fig:DIFFDIFF}
%\end{figure}

%\\
%\\
%\\
%From these representations, we can see that the effect of the cavity tends to accumulate electron density around electronegative atoms, such as Oxygen and Nitrogen, and on the aromatic rings of the nucleotide bases, while it removes it from electropositive elements, in accordance with previous observations made in the literature. This effect is thought to be linked to the minimization of the self-dipole energy term.

From the figure we observe that the field tends to localize the charge on the electron-rich regions of the system. The same trend was already observed in previous studies.~\cite{haugland2020coupled,haugland2021intermolecular, flick2018ab}  

% Before concluding our analysis it is important to stress that the long-wavelength approximation used in our calculation, despite being widely used by the polaritonic community,~\cite{kamper2023theory,svendsen2025effective,rokaj2018light} it is not optimal when the system is coupled to plasmonic cavities (as those investigated in our case). In this situation the device' size and the field wavelength are comparable with the molecular dimension. In this case a full minimal coupling treatment of the problem should be used.
The calculations presented so far have been performed in a dipole approximation framework. This limit is valid only if the molecular dimension is significantly smaller than the size of the device.
This is clearly not the case for a plasmonic setup.
In order to address the accuracy of our results
we compared the data presented above with those obtained by two modes ($+k$ and $-k$ with the same resonance frequency) minimal coupling calculations which include the exact exponential spatial dependence of the field.
This analysis has been performed using an alternative implementation of the QED-CC approach discussed in Ref.~\citenum{riso2023strong}. The resulting tests are presented in Fig.~S3 of the Supporting Information.
For these systems significant deviations from the discussed trends have not been observed. 
In particular minimal coupling effects generate, for the AT dimer, a slightly larger stabilization (compared to the dipole approximation case) of about 20 meV that anyway does not change the general trends observed above.

\paragraph{Conclusions.}

%Using the state of the art QED-CC model contained in $e^{\mathcal{T}}$~\cite{folkestad20201} , we approached a problem which already was studied by two other groups~\cite{tao2025probing,hou2024vibrational} . This problem was how the h-bond energy varies when a quantum field is applied to the AT and CG dimers. Unlike the former groups -who used an experimental approach and studied collective effects on vibrational levels-, we studied the problem by focusing on a theoretical approach and the effects on the single dimer's electronic energy levels. 

In this paper we proposed a new approach based on plasmonic cavities to perform DNA engineering based on quantum fields. 
The analysis based on QED-CC calculations demonstrated that depending on the field polarization/DNA strand orientation the nucleotides binding energy can be locally manipulated. 
In particular while for the AT dimer an hydrogen bond stabilization is observed for every allowed orientation, for the CG dimer case different orientations of the hydrogen bonds with respect to the field polarization direction can induce stabilization or destabilization of the bond.

This observation brings us to the conclusion that the combination of plasmonic cavities and DNA-origami-based mechanisms ~\cite{kosuri2019rotation,pumm2022dna,bryant2003structural,lebel2014gold} able to rotate and immobilize the DNA strand in a specific position might be a good strategy to maximize the cavity-induced effect and to select the orientations at which hydrogen bonds destabilization can be obtained.

In the AT case, instead a reverse approach might be used. In this situation, where only stabilizations are observed, we can imagine to use two properly displaced plasmonic cavities (as shown in Fig.~S4) to stabilize the neighboring nucleotides with consequent destabilization of the selected AT dimer.       
%\begin{figure}[ht!]
%    \centering
%    \includegraphics[width=\linewidth]{Images/Plasmonic_Cav_2(1).jpeg}
%    \caption{Representation of the double cavity system}
%    \label{fig:double_cav}
%\end{figure}

Despite promising, the observed effects are quite small (never exceed 5\% of the total binding energy), at least for the coupling values ($\lambda=0.05$ a.u.) used in this work. 
However, it is important to point out that coupling values comparable to the one used in our study can be easily obtained in plasmonic cavities also for a single molecule.
Indeed, in this kind of devices, often ultra-strong or even deep-ultra-strong coupling conditions can be reached ~\cite{hugall2018plasmonic,ameling2013microcavity,chikkaraddy2016single,lee2023strong,baranov2020ultrastrong} . In those limiting cases much larger relative effects should be observed.

\begin{acknowledgement}

The authors are thankful to Rosario R. Riso for insightful discussions and for the help with the minimal coupling QED-CC calculations.
R.A. and E.R. acknowledge funding from the
European Research Council (ERC) under the European
Union’s Horizon Europe Research and Innovation Programme
(Grant n. ERC-StG-2021-101040197 - QED-SPIN).

\end{acknowledgement}

\paragraph{Data Availability Statement}
The e$^{\mathcal{T}}$ program used to perform QED-CCSD calculations in this work is available at the main software's GitLab repository: \url{https://gitlab.com/eT-program/eT}.
The e$^{\mathcal{T}}$ code used to perform QED-CCSD calculations with the minimal coupling Hamiltonian can be found at the Zenodo of the original paper (\url{10.5281/zenodo.7035887}).~\cite{riso2023strong}

Examples input files, molecular geometries and instructions to reproduce the calculations are available at the following Zenodo link: \url{10.5281/zenodo.11977434}.

\begin{suppinfo}
The Supporting Information contains the geometries of the adenine-thymine and cytosine-guanine dimers, the plot of absolute energies polarization dependence, the plot of cavity induced charge density modifications and the study of the field spatial dependence effects on the hydrogen bonds dissociation potential energy surfaces.

\end{suppinfo}

%\bibliography{achemso}

\providecommand{\latin}[1]{#1}
\makeatletter
\providecommand{\doi}
  {\begingroup\let\do\@makeother\dospecials
  \catcode`\{=1 \catcode`\}=2 \doi@aux}
\providecommand{\doi@aux}[1]{\endgroup\texttt{#1}}
\makeatother
\providecommand*\mcitethebibliography{\thebibliography}
\csname @ifundefined\endcsname{endmcitethebibliography}  {\let\endmcitethebibliography\endthebibliography}{}

\newpage

\includepdf[pages=-]{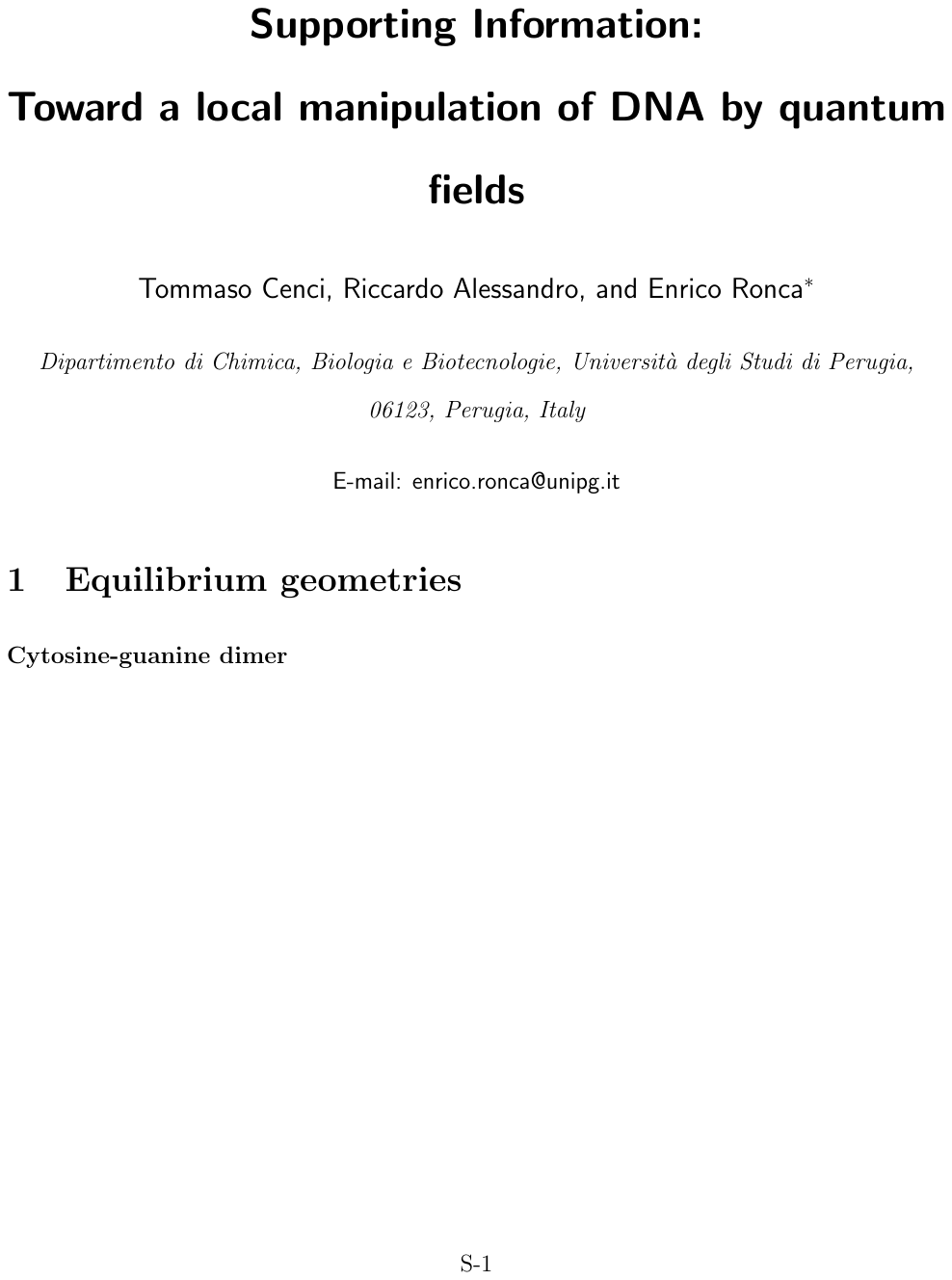}

\end{document}